# Topological acoustics in coupled nanocavity arrays

M. Esmann,[1] F.R. Lamberti,[1] A. Lemaître,[1] and N.D. Lanzillotti-Kimura[1]

[1] Centre de Nanosciences et de Nanotechnologies, CNRS, Université Paris-Sud, Université Paris-Saclay, C2N Marcoussis, 91460 Marcoussis, France

**Abstract**

The Su-Schrieffer-Heeger (SSH) model is likely the simplest one-dimensional concept to study non-trivial topological phases and topological excitations. Originally developed to explain the electric conductivity of polyacetylene, it has become a platform for the study of topological effects in electronics, photonics and ultra-cold atomic systems. Here, we propose an experimentally feasible implementation of the SSH model based on coupled one-dimensional acoustic nanoresonators working in the GHz-THz range. In this simulator it is possible to implement different signs in the nearest neighbor interaction terms, showing full tunability of all parameters in the SSH model. Based on this concept we construct topological transition points generating nanophononic edge and interface states and propose an easy scheme to experimentally probe their spatial complex amplitude distribution directly by well-established optical pump-probe techniques.

**Introduction**

The polymer trans-polyacetylene consists of a carbon chain with alternating C-C single and double bonds. Despite this simple chemical structure, trans-polyacetylene triggered numerous fundamental questions when it was discovered that it becomes electrically conducting upon halogen doping [1,2]. The mechanism behind this extraordinary property is conceptually captured by the celebrated Su-Schrieffer-Heeger (SSH) model [3,4]. Essentially, it describes spin-polarized electrons on a one-dimensional lattice with staggered nearest-neighbor hopping in the tight-binding approximation. Thereby, the SSH model supports the formation of topological solitons [5,6].

The particular importance of the SSH model lies in its ability to provide a simple, yet prototypical example of topological phase transitions in one dimension. This has spurred the study of many classical and quantum mechanical systems beyond polymers, which also allow a description by the SSH or related two-dimensional models. For example, ultra-cold atomic quantum gasses in optical lattices [7] have been used for the direct measurements of Zak phases (the Berry phase in a one-dimensional periodic medium) [8], for the observation of topological edge states [9] and for topological charge pumping [10,11]. In one- and two-dimensional photonic crystals and waveguides [12–14] topological edge states have also been reported [15–19] allowing one-way optical transport [20,21] if time reversal symmetry is broken. Very recently, the SSH model was implemented in a polaritonic system, supporting lasing edge modes that are robust to disorder [22] and in macroscopic acoustics [23,24]. Reports on studies of topological effects in nanoacoustics however remain sparse.

High frequency acoustic phonons, i.e., vibrations in solids in the GHz-THz range, are relevant in the determination of thermal and electronic transport properties and constitute a main source of decoherence in solid state quantum systems [25,26]. Acoustic phonons also represent a versatile platform for the study of wave dynamics and localized excitations featuring two main advantages with respect to the electronic and optic counterparts: First, due to their slow speed of propagation compared to light, acoustic phonons at 100 GHz-THz frequencies have wavelengths as short as a few nanometers. This allows the experimental implementation of very large systems [27], which may be considered infinite for all practical means [28]. Second, these frequencies and slow speeds render the resulting wave function quasi-static when probed optically, e.g. by pump-probe spectroscopy giving full access to complex amplitudes and phase information [29]. The fundamental building blocks to confine, shape and guide phonons at the nanoscale are finite size superlattices employed as distributed Bragg reflectors (DBRs) and phononic Fabry-Perot cavities formed by two DBRs enclosing a resonant spacer layer and coupled cavity arrays [30–32]. Although these versatile elements are well-established devices in nanophononics, their application to the study of topological order has remained greatly unexplored [28,33].

Here we propose to implement the SSH model in acoustic nanocavity arrays working in the 300 GHz-range with each cavity representing an atom along a one-dimensional chain of carbon. We demonstrate that our implementation supports the formation of topologically protected edge and interface states. We furthermore show that it is feasible to optically address the resulting topological acoustic states of the system with clear, easy to detect signatures by well-established pump-probe experiments. We demonstrate that it is furthermore possible to experimentally detect a sign reversal of the staggered hopping terms, which allows distinguishing fully symmetric and anti-symmetric edge states. Our results thus present a critical step forward towards the observation of more complex topology-driven physical effects in nanophononics with large potential to perform the optimization of interaction with electrons and photons based on the topological engineering of the structures.

**Results and Discussion**

The order of the single/double bonds in polyacetylene leads to two energetically degenerate isomers as depicted in the top panels of Figure 1a and b. Each of the $sp^2$-hybridized carbon atoms contributes a delocalized $\pi$-electron subject to the staggered hopping potential across double and single carbon bonds. For a chain of $N$ diatomic unit cells, this situation is captured via the second-quantized single-particle Hamiltonian [3,4,8,9,34]

$$\hat{H} = \sum_{n=1}^{N} v\left(a_n^+ b_n + h.c.\right) + w\left(a_n^+ b_{n-1} + h.c.\right) \quad (1)$$

Here, $v$ and $w$ are the inter- and intra-cell hopping integrals. The operators $a_n^{(+)}$ and $b_n^{(+)}$ annihilate (create) a particle in the $n$-th unit cell on sublattice a or b, respectively.

The two isomers in Figure 1a and b represent two topologically different phases P$_1$ and P$_2$ [34,35]. Essentially, if the intra-cell hopping dominates ($v < w$), the band structure of an infinite P$_2$ type lattice is characterized by a non-trivial winding number $g = 1$. Equivalently, a particle moving adiabatically on a closed path in momentum space across the first Brillouin zone acquires a Zak phase [8,34–36] (the analog of a geometric Berry phase [37] for Bloch bands) of $\phi_{Zak} = \pi$. In contrast, for the topologically trivial phase P$_1$ intercell hopping dominates ($v > w$) and both $g$ and $\phi_{Zak}$ are zero. A central prediction of the SSH model is the formation of interface states at topological transition points between the phases P$_1$ and P$_2$ [3,4] (see Figure 2) or between the non-trivial phase P$_2$ and vacuum [34], that is, when the polymer chain is terminated as shown in Figure 1b.

To establish an analog of the Hamiltonian (1) in nanophononics, we start from an individual nanophononic cavity, which is usually constructed from a resonant spacer layer of acoustic thickness $m\lambda/2$ ($m \in \mathbb{N}$, $\lambda$ acoustic wavelength) enclosed in between two Distributed Bragg Reflectors (DBRs). A DBR is a structure that presents a periodic modulation of the elastic properties and exhibits spectral intervals of high acoustic reflectivity. These stop-bands correspond to acoustic band gaps of the equivalent infinite periodic lattice. If choosing alternating layers of acoustic thickness $\lambda/4$, the first acoustic bandgap of the DBR is centered exactly at the resonance of the spacer and the structure acts as a Fabry-Perot resonator. Such a cavity can be understood as the phononic analog of a one-dimensional atom with two potential barriers enclosing a potential well [27]. Accordingly, its solutions to the acoustic wave equation form a spectrum of resonant modes, which are localized in the spacer layer.

A first extension of this concept is to couple two resonant cavity spacers by placing two DBRs around and one in between them. The system can then be regarded as the phononic analog of a diatomic molecule [38]. It exhibits resonant modes that represent symmetric and anti-symmetric linear combinations of the individual cavity resonances. For the implementation of the SSH model the key idea is to couple a chain of such phononic dimers. The necessary staggering of the hopping is introduced by using DBRs with different numbers of layers (and hence different reflectivities) for the inter- and intracell hopping. The resulting structures are schematically shown in the top panels of Figure 1c and d. Making the direct analogy to the chemical structure of polyacetylene, we identify each Fabry-Perot spacer (dark blue) with a site (a carbon atom) and each DBR with a C-C single or double bond.

As a first example, we simulated the topological phases P$_1$ and P$_2$ and demonstrate the emergence of a topological edge state for the non-trivial phase. Based on the acoustic properties of GaAs and AlAs we designed a structure resonant at $f_0 = 300\,\text{GHz}$, i.e. for an acoustic wavelength of $\lambda = c_{GaAs}/f_0 = 16\,\text{nm}$ ($c_{GaAs}$ speed of sound in GaAs) in the GaAs spacers of the cavity chain. DBRs were simulated with thicknesses of a quarter wavelength per layer such that the first high reflectivity stop band is centered around $f_0$. In order to establish $v < w$, i.e. the non-trivial phase P$_2$, we used DBRs of 8.5 layer pairs inside unit cells (16 layers GaAs and 17 layers of AlAs) and DBRs of 4.5 layer pairs to connect adjacent unit cells. In Figure 1c and d the result of a corresponding transfer-matrix calculation on a lattice of 6 diatomic unit cells is shown for both P$_2$ and P$_1$. Here, we plotted the mechanical strain distribution in the nanophononic cavity chain at $f_0$. We indeed observe the formation of a confined edge state at $f_0$ in panel d showing the topologically nontrivial phase P$_2$. In agreement with the predictions of the SSH model [35] we also find that the mode selectively populates only sublattice b. That is, the strain exhibits anti-nodes on each spacer layer of sublattice b and nodes on all sites of sublattice a. Furthermore, subsequent antinodes of the strain alternate in sign. The overall envelope of the edge state decays evanescently into the structure. In contrast, we find no such state for the topologically trivial phase P$_1$ (panel c). The only resonances in this case are Bloch-like modes of the finite system, which occupy both sublattices.

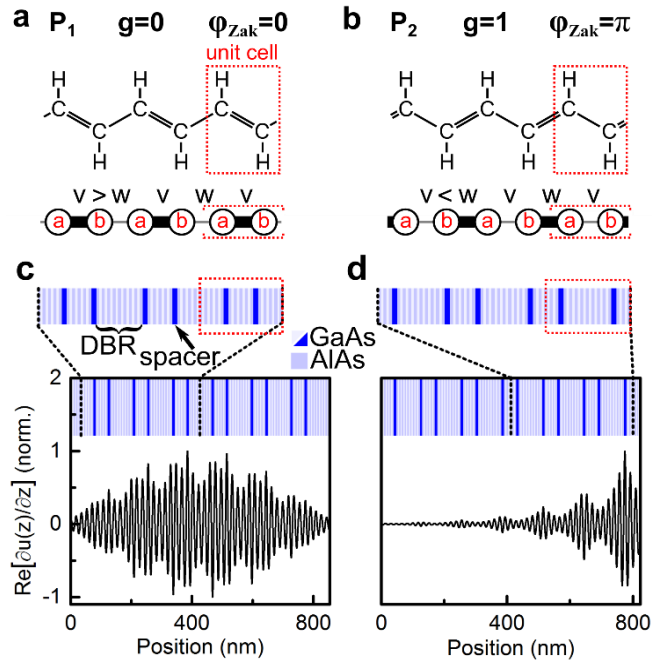

**Figure 1 SSH model in an acoustic nanocavity array, a-b,** Topological phases P$_1$ (trivial) and P$_2$ (non-trivial) of the SSH model with corresponding carbon chains of polyacetylene. **c-d,** (top) Phases P$_1$ and P$_2$ implemented in a coupled nanophononic cavity array with resonant spacer layers representing atomic sites and thick and thin distributed Bragg reflectors (DBRs) representing weak and strong hopping links, respectively. (bottom) Strain distributions $\partial u(z)/\partial z$ in nanophononic cavity arrays of six unit cells for a Bloch-like mode of the trivial phase (c) and the topological edge mode of the non-trivial phase at $f = 300\,\text{GHz}$ (d).

The next extension of our implementation is to investigate the presence of topological interface states when concatenating the two topological phases P$_1$ and P$_2$ as presented in Figure 2. In panels a and b the two possible configurations of an interface are sketched for polyacetylene and for a nanophononic cavity array. One can either construct a monomer-like defect (panel a) represented by one spacer layer connected with two weak hopping links (two wider DBRs) or a trimer-like defect composed of three sites connected by strong hopping links (panel b). Figure 2c and d show the results of corresponding model calculations, again using DBRs of 4.5 and 8.5 layer pairs, respectively, and a design frequency of $f_0 = 300\,\text{GHz}$. Both the monomer and the trimer state are resonant at $f_0$ and their strain distributions with exponentially decaying envelopes are spatially localized around the defects in the corresponding structures. As already seen for the edge state in Figure 1d, the strain distribution of the monomer state exhibits antinodes of alternating sign on the central defect and on every second other spacer. Antinodes are found on the sites in between. For the trimer-like case the situation is different: On the central site of the defect the state exhibits a node, whereas antinodes of opposite sign but equal magnitude are observed on the two outer spacer layers of the defect. In the remainder of the structure, the same behavior as for the monomer is found with antinodes in strain on every second other site and nodes in between. Again, these findings are in full agreement with the predictions of the generalized SSH model [35].

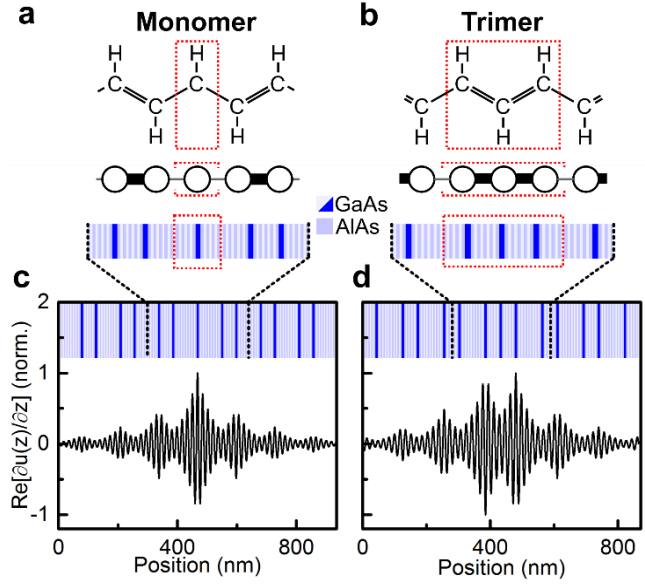

**Figure 2 Mono- and Trimer-like SSH interface states in nanophononic cavity array; a,** Monomer defect in polyacetylene and coupled phononic cavities. **b,** Trimer defect **c-d,** Strain distribution of topological monomer (c) and trimer defect states (d) in coupled phononic cavities at $f_0 = 300\,\text{GHz}$ superimposed with the GaAs/AlAs layer structure. DBRs contain 4.5 and 8.5 layer pairs, respectively.

Implementing hopping terms with DBRs not only allows tuning the magnitude but also the phase of the hopping integral. In particular, the sign of the hopping can be reversed by introducing an additional hopping phase of $\pi$ as sketched in the insets of Figure 3. Using again the topological edge state of the phase P$_2$ as an example, we discuss how a sign reversal of the hopping can be achieved with nanophononic cavities and how the edge state is affected. First, consider again the structure composed of sites separated by DBRs of 4.5 and 8.5 layer pairs (Figure 3a). The phase which is acquired by a phonon upon propagating between the centers of adjacent lattice sites on sublattice b is $\phi_{AS} = \phi_v + \phi_w + 2\pi$. Here, $\phi_v$ and $\phi_w$ represent the propagation phases through the two DBRs and the additional phase of $2\pi$ accounts for the propagation through the spacer layers. Since the DBRs in our example are constructed from alternating layers of acoustic thickness $\lambda/4$, a DBR with $m$ layers leads to a propagation phase of $\phi = m \cdot \pi / 2$. In total the phonon therefore picks up a phase of $\phi_{AS} = 7.5\pi$, which translates into a state with alternating sign of the strain on subsequent sites of sublattice b. We refer to this configuration as an anti-symmetric wavefunction.

In Figure 3b, we repeated the calculation of the edge state but added one extra pair of material layers to each DBR of type $v$ (i.e. 9.5 instead of 8.5). The overall propagation phase between neighboring sites on sublattice b hence changes to $\phi_S = 8\pi$. As clearly visible from the strain pattern, this drastically changes the symmetry of the state with all nodes of the strain on sublattice b now oscillating in phase. We hence term this configuration a symmetric wavefunction. We note that in principle the additional propagation phase, which we have introduced, can be gauged away in the SSH model by incorporating it into the basis states [9,35]. Thus, all considerations above concerning topological phases and winding numbers still hold. Nevertheless, as we will see shortly, the change in symmetry of the states has a significant impact on the experimental signatures of the topological states in optical detection.

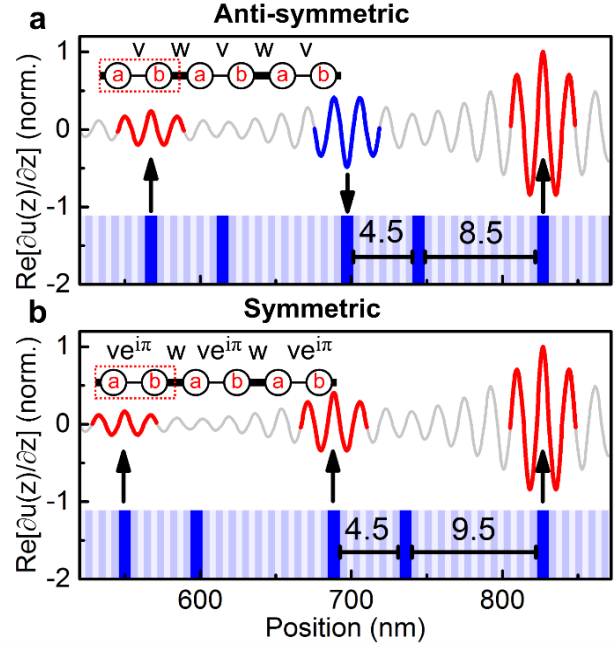

**Figure 3 Symmetric and Anti-symmetric topological edge states in nanophononic cavity array; a,** Anti-symmetric edge state on sublattice b with alternating sign of the strain (red and blue), and layer-structure with alternating DBRs of 4.5 and 8.5 layer pairs. **b,** Symmetric edge state. By the addition of one layer pair in the thicker DBRs the hopping integral acquires an additional phase of $\pi$.

In the last part of this paper we discuss the experimental feasibility of detecting the dynamics in a particular acoustic atom in a coupled acoustic nanocavity array by all-optical means. To this end, we simulated near-infrared pump-probe microscopy measurements following the approach presented in Refs. [39–41].

The coherent generation and detection of acoustic phonons involves the use of ultrafast pulsed lasers to create and measure coherent mechanical oscillations in the GHz/THz range. In this technique, a first ultrafast laser pulse interacts with the sample, generating coherent acoustic phonons. The coherent vibrations induce changes in the optical reflectivity of the sample. A second delayed pulse probes the instantaneous optical reflectivity of the sample. By changing the delay between the pump and probe pulses it is possible to reconstruct the time-resolved reflectivity of the sample.

For practical purposes, generation and detection of coherent phonons can be considered as two independent experiments. For instance, the generation can be localized in a metallic thin film at the sample surface, generating a broadband acoustic pulse that interacts with the semiconductor sample below, whereas the detection takes place inside the semiconductor structure itself. Two mechanisms mediate the detection process, the displacement of interfaces between layers (i.e. changes in the shape and size of the sample), and the photoelastic effect (i.e. changes in the dielectric function of the acoustic materials induced by phonon strain). When the laser wavelength is close to any of the electronic transitions in the materials forming the sample, the contribution of the displacement of the interfaces, however, becomes negligible. Moreover, the photoelastic effect is strongly resonant. By tracing out the optical reflectivity in time upon systematic variation of the temporal delay between pump and probe pulse, the dynamic evolution of the phonons generated by the pump pulse are reconstructed.

To simulate the experiment described above, we consider a broadband, Fourier limited phonon pulse generated outside the topological nanophononic structure, e.g. by employing a thin metallic layer as a transducer [31]. For the amplitude spectrum $g(\omega)$ of the pulse we choose a Gaussian centered at $f_0 = 300\,\text{GHz}$ with a full width of $2\Delta f = 40\,\text{GHz}$:

$$g(\omega) \propto \frac{1}{\sqrt{2\pi}\Delta\omega} \exp\left(-\frac{(\omega-\omega_0)^2}{2\Delta\omega^2}\right) \tag{2}$$

The time-resolved detection process of changes in optical reflectivity $\Delta R(t)$ is modeled by calculating the photoelastic overlap integral [39]

$$\Delta R(t) \propto \int_0^L p(z) \frac{\partial u(z,t)}{\partial z} E_p^2(z) dz \tag{3}$$

Here, $L$ is the length of the phononic structure, $\partial u(z,t)/\partial z$ the strain of the propagating phonon pulse and $E_p(z)$ the electric field distribution of the probe pulse. For the electric field of the probe pulse we use a standing wave pattern [39], which can for example be generated by an optical DBR placed between the acoustic nanocavity array and the substrate. $p(z)$ is the material-dependent photoelastic constant. If the spectrum of the probe pulse covers an electronic transition in the system, $p(z)$ exhibits a resonance, too. This enables localized detection schemes [40,41]: First, considering that the typical layer thicknesses in the nanophononic structures proposed here are only on the order of 10 nm, each layer acts as a quantum well with a thickness-dependent exciton energy. Hence, the detection process can be completely dominated by the thickest layers in the system, in our case by the GaAs spacer layers. Moreover, GaAs layers can be doped with Indium, inducing an additional local red-shift of the electronic transition without considerably modifying the elastic behavior of the layer. In this way, a *single* spacer inside a coupled cavity structure can act as a selective detector for coherent acoustic phonons.

Let us analyze two distinct cases of topological cavity arrays. The first one, described in the inset of Fig. 4a supports a symmetric monomer topological state, i.e. the wavefunction of the interface state presents maxima of equal (positive) sign on neighboring unit cells (indicated with red). The second array (inset of Fig 4b) also supports a topological monomer interface state, but the wavefunction maxima on neighboring unit cells are of opposite sign (indicated with blue and red). This staggered symmetry is achieved by adding one pair of material layers to every other DBR following the concept shown in Fig. 3. In Fig. 4a and b (left panels) we show simulated time traces when the central (first neighbor) spacer is selectively addressed (indicated by black and green dots) by setting $p(z)$ to a constant inside the spacer and zero everywhere else. In both cases, we observe an exponentially decaying envelope, related to the lifetime of the phonons in the cavity array. The signal is generated by both confined and propagating phonon modes that are contained in the incident phonon spectrum. The spectral components of the pulse, which do not match any localized eigenstate of the structure, decay at relatively short times. At sufficiently large time scales ($\tau > 2\,\text{ns}$), we hence observe a completely periodic modulation of transient reflectivity which is completely dominated by the topological interface state.

By selectively probing spacers in two consecutive unit cells of the symmetric structure (panel a) we observe in-phase transient reflectivity traces (see zoomed-in right panel). Conversely, when studying the structure with anti-symmetric spatial wavefunction profiles (panel b), the time traces are in counter phase. Comparing the time traces hence directly allows to measure the symmetry properties of an acoustic topological interface state in the coupled-cavity structure. By subsequently addressing each one of the acoustic atoms separately, the dynamics of the system becomes fully accessible.

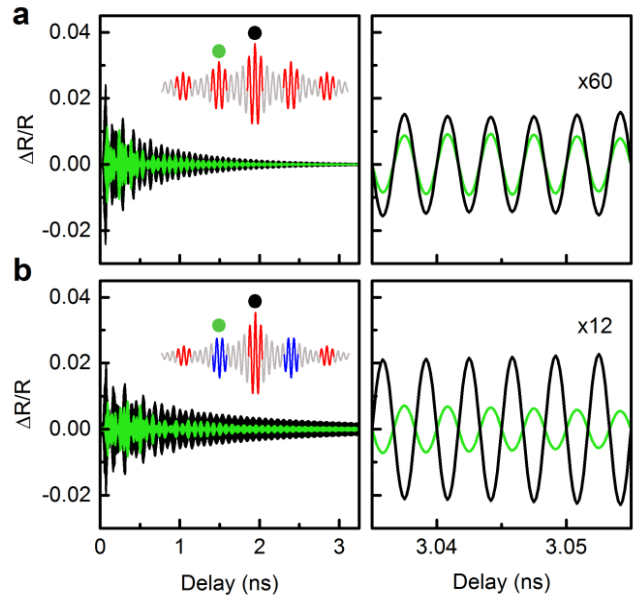

**Figure 4** Simulated transient reflectivity traces of topological defect states in nanophononic cavity arrays; **a,** (left) Symmetric topological monomer defect state at 300 GHz (3.5 and 6.5 layer pairs per DBR). Time traces are simulated for photoelastic contributions from the central (black) and left-of-center (green) spacer layer. (right) Zoom-in of the time traces showing in-phase modulation of both cavities' signatures. **b,** (left) Anti-symmetric topological monomer defect state (3.5 and 7.5 layer pairs per DBR). (right) Zoom-in of the time traces presenting out-of-phase modulation indicating out-of-phase oscillation of the strain in the central and left-of-center cavity. For better visibility, vertical scales in the right panels of a and b are magnified by factors of 60 and 12, respectively.

**Conclusions and Outlook**

In spite of its basic character, the SSH model keeps unveiling exciting physics. In this work, we presented an implementation based on coupled acoustic nanocavities. In these structures, each nanocavity plays the role of a carbon atom and the DBRs determine the complex coupling constants between them.

We analyzed the case of typically ten coupled cavities, where by tuning the reflectivity of the mirrors we were able to control the topological winding numbers of the structures. The simplicity of this mapping allowed us to easily study key cases as the confinement in monomer and trimer states, and symmetric and anti-symmetric acoustic states. The proposed toolbox is compatible with optical probes such as Raman scattering and picosecond ultrasonic coherent phonon sensing.

Coherent pump-probe experiments are not only a sensitive tool for the detection of confined acoustic states at topological transition points, but also give valuable insight into their symmetry properties. The temporal phase in different acoustic atoms could be accessed by selectively detecting the displacement along the acoustic atomic chain. In other words, we presented a simple structure able to mimic the topological physics of polyacetylene and we proposed a standard optical experiment that can probe the full resulting phonon wave function. These results represent a first step in the simulation of more complex Hamiltonians. For instance, the implementation of a time reversal symmetry breaking scheme, the use of active phononic materials, exploring anharmonicity effects or tunable and reconfigurable phononic systems in topological arrays of coupled cavities are just a few examples of exciting perspectives of this work.

**Acknowledgements**

The authors acknowledge funding through the ERC Starting Grant No. 715939 Nanophennec, and through a public grant overseen by the French National Research Agency (ANR) as part of the "Investissements d'Avenir" program (Labex NanoSaclay, reference: ANR-10-LABX- 0035).